\documentclass[sigconf]{acmart}
\usepackage{fontawesome5}
\usepackage{multicol}
\usepackage{enumitem}
\AtBeginDocument{%
  \providecommand\BibTeX{{%
    \normalfont B\kern-0.5em{\scshape i\kern-0.25em b}\kern-0.8em\TeX}}}
\setcopyright{acmcopyright}
\copyrightyear{2025}
\acmYear{2025}
\setcopyright{rightsretained}
\acmConference[UIST Adjunct '25]{The 38th Annual ACM Symposium on User Interface Software and Technology}{September 28-October 1, 2025}{Busan, Republic of Korea}
\acmBooktitle{The 38th Annual ACM Symposium on User Interface Software and Technology (UIST Adjunct '25), September 28-October 1, 2025, Busan, Republic of Korea}\acmDOI{10.1145/3746058.3758469}
\acmISBN{979-8-4007-2036-9/2025/09}
\begin{document}
\renewcommand{\thefootnote}{\fnsymbol{footnote}}
\newcommand{\todo}[1]{{\color{red} #1}}

\title{Facilitating Longitudinal Interaction Studies of AI Systems}
\author{Tao Long}
\affiliation{
  \institution{Columbia University}
  \city{New York}
  \state{New York}
  \country{USA}
}
\email{long@cs.columbia.edu}

\author{Sitong Wang}
\affiliation{
  \institution{Columbia University}
  \city{New York}
  \state{New York}
  \country{USA}
}
\email{sitong@cs.columbia.edu}

\author{Émilie Fabre}
\affiliation{
  \institution{The University of Tokyo}
  \city{Tokyo}
  \country{Japan}
}
\email{fabre@g.ecc.u-tokyo.ac.jp}

\author{Tony Wang}
\affiliation{
  \institution{Cornell University}
  \city{Ithaca}
  \state{New York}
  \country{USA}
}
\email{yw2567@cornell.edu}

\author{Anup Sathya}
\affiliation{
  \institution{University of Chicago}
  \city{Chicago}
  \state{Illinois}
  \country{USA}
}
\email{anups@uchicago.edu}

\author{Jason Wu}
\affiliation{
  \institution{Apple}
  \city{Seattle}
  \state{Washington}
  \country{USA}
}
\email{jason_wu8@apple.com}

\author{Savvas Petridis}
\affiliation{
  \institution{Google DeepMind}
  \city{New York}
  \state{New York}
  \country{USA}
}
\email{petridis@google.com}

\author{Dingzeyu Li}
\affiliation{
  \institution{Adobe Research}
  \city{Seattle}
  \state{Washington}
  \country{USA}
}
\email{dinli@adobe.com}

\author{Tuhin Chakrabarty}
\affiliation{
  \institution{Salesforce AI Research}
  \city{New York}
  \state{New York}
  \country{USA}
}
\email{tchakrabarty@cs.stonybrook.edu}

\author{Yue Jiang}
\affiliation{
  \institution{Aalto University, University of Utah}
  \city{Espoo}
  \country{Finland}
}
\email{yue.jiang@aalto.fi}

\author{Jingyi Li}
\affiliation{
  \institution{Pomona College}
  \city{Claremont}
  \state{California}
  \country{USA}
}
\email{jingyi.li@pomona.edu}

\author{Tiffany Tseng}
\affiliation{
  \institution{Barnard College}
  \city{New York}
  \state{New York}
  \country{USA}
}
\email{ttseng@barnard.edu}

\author{Ken Nakagaki}
\affiliation{
  \institution{University of Chicago}
  \city{Chicago}
  \state{Illinois}
  \country{USA}
}
\email{knakagaki@uchicago.edu}

\author{Qian Yang}
\affiliation{
  \institution{Cornell University}
  \city{Ithaca}
  \state{New York}
  \country{USA}
}
\email{qianyang@cornell.edu}

\author{Nikolas Martelaro}
\affiliation{
  \institution{Carnegie Mellon University}
  \city{Pittsburgh}
  \state{Pennsylvania}
  \country{USA}
}
\email{nikmart@cmu.edu}

\author{Jeffrey V. Nickerson}
\affiliation{
  \institution{Stevens Institute of Technology}
  \city{Hoboken}
  \state{New Jersey}
  \country{USA}
}
\email{jnickers@stevens.edu}

\author{Lydia B. Chilton}
\affiliation{
  \institution{Columbia University}
  \city{New York}
  \state{New York}
  \country{USA}
}
\email{chilton@cs.columbia.edu}


\renewcommand{\shortauthors}{Facilitating Longitudinal Interaction Studies of AI Systems}

\begin{abstract}
UIST researchers develop tools to address user challenges. However, user interactions with AI evolve over time through learning, adaptation, and repurposing—making one-time evaluations insufficient. Capturing these dynamics requires longer-term studies, but challenges in deployment, evaluation design, and data collection have made such longitudinal research difficult to implement. Our workshop aims to tackle these challenges and prepare researchers with practical strategies for longitudinal studies. The workshop includes a keynote, panel discussions, and interactive breakout groups for discussion and hands-on protocol design and tool prototyping sessions. We seek to foster a community around longitudinal system research and promote it as a more embraced method for designing, building, and evaluating UIST tools.\\
 \href{https://longitudinal-workshop.github.io/}{\faLink[regular]\ : \texttt{\textbf{https://longitudinal-workshop.github.io/}}}
\end{abstract}

\maketitle

\section{Workshop Motivation and Goal}

UIST researchers have been developing tools to address user challenges. Recent advances in AI foundation models have made this even easier—capabilities in code generation, synthesis, and accessible APIs have significantly lowered the cost and barriers to turning ideas into functional AI prototypes and deploying them quickly.

At the same time, the way users interact with AI tools is evolving rapidly. People's knowledge, attitudes, and behaviors around AI shift as they gain experience over time—through trial and error, social learning, and by customizing or repurposing tools. This makes user interactions more complex and harder to measure using traditional short-term evaluations.
As new opportunities and complexities emerge from both the system and user sides, one fixed system or one-time evaluation is no longer sufficient to capture evolving practices. There is a clear need to rethink how we study interaction and to examine these dynamics with greater fidelity.

Longitudinal studies are uniquely timely and necessary, as they offer deep insights and directly address these shifts. AI tools particularly benefit from such studies due to their dynamic nature and the ease with which they can be updated. New interaction mechanisms can also be discovered through richer user data. 

Many UIST papers now focus on technical evaluations or isolated user studies as forms of evaluation. While these approaches are effective for assessing tools in single-session contexts, they may overlook real-world implications and the evolving nature of user interactions with these tools~\cite{taolongitudinal}. Longitudinal studies allow researchers to go beyond initial proof-of-concept work and understand how the tools perform over time in real-world conditions~\cite{10.1145/2212776.2212706}. This creates a stronger foundation for both technical contributions and industry relevance—offering deeper insights into usability, appropriation, sustained use, and workflow integration.

Specifically, longitudinal interaction studies allow researchers to better understand how users’ experiences and perceptions with AI tools evolve over time, generating rich interaction data that helps build more effective, trusted, and user-aligned systems~\cite{nielsen_longitudinal_2021, wang2025role, 10.1145/3313831.3376301, taolongitudinal,10.1145/3706599.3720135}. For instance, such studies enable the analysis of several key aspects of technology adoption and long-term use, offering a more accurate understanding of real-world impact and behavioral change.
It captures \textit{(i) novelty and placebo effects}, where initial excitement or belief in the tools may skew perceptions before stabilizing~\cite{lai2017literature, 9288720, fitbit_novelty, koch2018novelty, long9, novelty_microsoft, placebo}; tracks \textit{(ii) learning phases} as users explore and build mental models of the system~\cite{mira, riyatraining,long3, long7, moran_everyday_2002, berker2005domestication}; examines \textit{(iii) customization and personalization}, as users tailor tools to their needs and behaviors~\cite{Muller_Neureiter_Verdezoto_Krischkowsky_AlZubaidi-Polli_Tscheligi_2016, 10.1145/2660398.2660427,10.1145/2157689.2157804}; investigates \textit{(iv) appropriation}, where users repurpose tools for unintended tasks~\cite{DesigningForAppropriationAlanDix, dourish_appropriation_2003, mullerappro1, mullerappro2, carrollappropriation,10.1145/3643834.3661613}; and observes \textit{(v) shifts in usage patterns}, such as evolving user behaviors around tool features, including practices like multiuser collaboration and the use of different types of memory within agentic AI systems.~\cite{feedquac,10.1145/2598510.2598540, ReelFramer, 10.1145/3544548.3581115,long_challenges_2023, disco, nielsen1992finding}. It considers \textit{(vi) shifts in perception and capability}, including how users’ views of the task and system evolve, along with their changing mental models and skill levels~\cite{liAudienceImpressionsNarrative2025, 10.1145/3613904.3642187, socialdynamics, Ross, long7, Tweetorial_ICCC, longitudinalReview}. Finally, it explores \textit{(vii) (un)sustained use and workflow integration}, as users either adopt, partially adopt, or abandon the tool, and potentially change or integrate it into broader daily workflows~\cite{10.1145/3613904.3642505, 10.1145/3428361.3428469, jumpstarter,taolongitudinal,liuDigitalDoubleHatters2025}.

All these aspects demonstrate how longitudinal studies provide deeper, more comprehensive insights into developing emerging technologies and novel interactions, aligning with UIST’s focus while connecting to industry perspectives. This moment—when AI tool development and user evolution are rapidly accelerating—is ideal for the UIST community to invest in longitudinal methods that bridge strong proof-of-concept prototypes with real-world, sustained deployment.

Despite these benefits, 
researchers often face significant challenges in implementing longitudinal studies~\cite{ludlow2011design}. The complexities associated with long-term tool deployment and evaluation—such as system bug fixes, recruitment, retention
, protocol and metric selection, and the ability to study interactions without constant observation—have hindered deeper exploration of the longitudinal interaction between users and tools~\cite{soprano2024longitudinal, doi:10.1177/1609406920917493, abbad2016analyzing, thomson2003hindsight, ludlow2011design}.

Therefore, our workshop aims to foster discussion that facilitates longitudinal research in HCI and UIST, with three specific goals:

\begin{enumerate}[left=0px]

    \item \textbf{Identify existing challenges in conducting longitudinal UIST research and propose solutions.}
Longitudinal research in UIST is full of potential—but also fraught with practical, methodological, and conceptual challenges. Key issues include:
\textbf{(1a)} Designing effective protocols that balance check-in frequency, data richness, and participant burden while ensuring in-the-wild authenticity and ethical integrity;
\textbf{(1b)} Building robust, adaptable systems that can endure bugs, updates, and shifting user needs over time, while also supporting customization, transparency, and data logging or recording practices that respect user privacy;
\textbf{(1c)} Clarifying longitudinal contributions to UIST by articulating what unique insights and claims this method enables, and how it advances knowledge in ways short-term studies cannot;
\textbf{(1d)} Bridging academic and industry approaches, which differ in scale, timelines, and goals, to foster sustainable collaborations that enable shared tools, data, and methodologies;
and \textbf{(1e)} Embracing new frontiers with human-AI studies, where both users and AI agents evolve over time—raising questions about adaptation, personalization, and how to treat AI-as-participant in long-term deployments. After conference, the participants and organizers will compile a workshop summary\footnote{Similar to https://unstable.design/mutualbenefit/outcomes/} and post it on the website. 
    
    \vspace{5px}
    \item \textbf{Offer participants hands-on experience in designing longitudinal protocols and prototyping longitudinal systems.} Organizers with longitudinal study experience help participants overcome mental barriers, walk through simplified protocol and prototype design, preparing them for future studies.

    \vspace{5px}
    \item \textbf{Foster a community network of researchers interested in longitudinal research, and promote longitudinal as a more embraced method for designing, building, and evaluating future UIST tools.}
    \vspace{3px}
\end{enumerate}


\section{Workshop Schedule and Plan}


\subsubsection*{\textbf{Schedule}} We structure our one-day workshop into four phases. There is no participant selection process, in accordance with UIST's requirements. The finalized workshop schedule will be communicated to the registered participants via email and the workshop website: \href{https://longitudinal-workshop.github.io/}{\faLink[regular] \textbf{\texttt{longitudinal-workshop.github.io/}}}

\vspace{6px}
\noindent{\textbf{Phase 1 (9:00–11:00) — Welcome, Introduction, and Familiarizing participants with longitudinal practices}}

\textit{{(9:00-9:30) \underline{Opening Presentation, 30 min}}}  
Participants will be welcomed by the organizers with a presentation introducing key concepts, practices, and examples of longitudinal HCI research.

\textit{{(9:30-10:30) \underline{Participant Introductions, 60 min}}}
Each participant will give a short self-introduction, optionally referring to their submitted materials or project interests. This segment builds a shared understanding of participant backgrounds and goals.

\textit{{(10:30-11:00) \underline{Keynote, 20 min + 10 min Q\&A}}}  
The keynote speaker (TBD) is a leading expert in industry-based longitudinal research and will share recent developments in tools and products for long-term user evaluation.

\vspace{6px}
\noindent{\textbf{Phase 2 (11:00–12:00, 13:15–14:00) — Navigating the challenges of longitudinal HCI/UIST studies}}

\textit{{(11:00-12:00) \underline{Table Activity \#1: Identifying Challenges, 60 min}} }\\
Participants will be grouped into tables based on their interests and invited to share their previous work or potential new UIST interactive system ideas. Using a guided framework, they will collaboratively identify key challenges in conducting longitudinal studies. Work is done in pairs, followed by short table presentations.

\textit{{(13:15-14:00) \underline{Industry Panel, 45 min}}}
Following an optional group lunch for workshop attendees, a panel of product managers and user research experts from companies or startups will discuss how their work relates to longitudinal themes—like tracking user behavior over time, iterative development, and product adoption. This panel bridges industry and academic practices.

\vspace{6px}
\noindent{\textbf{Phase 3 (14:00–16:00) — Gaining hands-on experience in designing longitudinal protocols and systems}}

\textit{{(14:00-15:00) \underline{Table Activity \#2: Study Protocol Design, 60 min}}}\\
Participants will work in pairs to design a new longitudinal study protocol. This includes formulating research questions, selecting appropriate methodologies and variables, determining study duration, outlining recruitment strategies, and planning for pilot testing. Each table will be assigned a specific scenario constraint (e.g., multi-week feature engagement, short-term workflow appropriation). Later, each table will present their protocols, share points of uncertainty, and reflect on trade-offs or design tensions.

\textit{{(15:00-16:00) \underline{Table Activity \#3: System Prototyping, 60 min}}}\\
Participants will prototype an AI tool of their choice—whether it is a current project, a past one, or a system they are interested in—to support their proposed studies, including sketches of five different wireframes and the design of system logging structures (e.g., what data to collect, how, and when; how to support longitudinal continuity). Each table will present their prototypes and questions.

\vspace{6px}
\noindent{\textbf{Phase 4 (16:00–17:00) — Reflection, Summary, and Future}}

\textit{{(16:00–16:40) \underline{Table Activity \#4: Reflection, 40 min}}} 
Participants will document their protocol designs and system prototyping insights (Activities \#2 and \#3) related to the initially identified challenges (Activity \#1). This reflection helps trace the evolution of their ideas and begin articulating potential contributions or open questions. Organizers will also share and reflect on workshop outcomes and initiate the development of a framework or toolkit repository to foster collaboration, as outlined in the post-workshop plan.

\textit{{(16:40–17:00) \underline{Closing, 20 min}}} 
The session will conclude with a short keynote given by Lydia Chilton, group photo, post-workshop planning (e.g., Discord/mailing list setup), and acknowledgments.

\subsubsection*{\textbf{Expected size of attendance}} We anticipate 30 to 35 attendees to participate in our UIST workshop. We believe this is a suitable size for community building and active participation and discussion related to our planned interactive activities.

\subsubsection*{\textbf{Post-workshop plans}}

We will share results and grow the community by: (1) co-writing a blog post summarizing key insights and future research directions, published on the workshop website; (2) co-creating living GitHub artifacts, including an annotated, up-to-date SIGCHI longitudinal papers collection, jumpstart guidelines and protocol templates, and a potential comprehensive survey or methods paper on longitudinal HCI/UIST tools; and (3) maintaining engagement via a SIGCHI Discord subchannel and mailing list to foster discussion, share ideas, and organize networking events.




\begin{acks}
We gratefully thank Katy Gero, David Ledo, and Julien Porquet for their valuable feedback and assistance with this proposal.  
\end{acks}
\bibliographystyle{ACM-Reference-Format}
\bibliography{bib}

\appendix

\section{Organizers}

\small
\vspace{2px}

\textbf{Tao Long} (Columbia University) is a PhD student whose research focuses on understanding and developing human–AI experiences, with the goal of making AI-powered tools more usable, useful, and seamlessly integrated into everyday workflows. He has led multiple long-term deployment studies of multimodal AI tools in creative and productivity contexts, to better how how users initiate, internalize, adapt to, and integrate AI into their work.

\vspace{2px}

\textbf{Sitong Wang} (Columbia University) is a PhD student at Columbia University. Her current work focuses on developing AI-powered systems to enhance human creativity and productivity. She is also interested in studying how to integrate these tools into existing workflows and examining their long-term impact on everyday life and the workplace.

\vspace{2px}

\textbf{Émilie Fabre} (University of Tokyo) is a PhD candidate at the University of Tokyo. Advised by Jun Rekimoto and Yuta Itoh, her main work focuses on enabling virtual entities in XR to better interact and communicate with our reality. Her strong technical background led her to bridge XR, robotics, and AI with the goal of creating corporeal virtual agents for general-purpose applications. With an emphasis on HCI and HRI, she strives to make interactions with virtual agents feel intuitive, expressive, and grounded in the physical world.

\vspace{2px}

\textbf{Tony Wang} (Cornell University) is a PhD student at Cornell University. His research explores novel approaches to designing and building AI systems that support multistakeholder communication and collaboration in domains such as mental health, online communities, and computer-supported cooperative work. Most recently, his interests have been focused on designing, developing, and deploying novel longitudinal reinforcement learning systems in the mental health domain.

\vspace{2px}

\textbf{Anup Sathya} (University of Chicago) is a PhD student at the University of Chicago, advised by Ken Nakagaki. His research explores how tangible and situated design can encourage more intentional technology use and support wellbeing. He focuses on creating household objects and systems that introduce purposeful friction to reduce technology overuse. His work has received multiple design awards and has been featured in major media outlets including CBC Radio and 404 Media.

\vspace{2px}

\textbf{Jason Wu} (Apple) is a Research Scientist at Apple in the Human-Centered Machine Intelligence group. Previously, he received a PhD in HCI from Carnegie Mellon. In his research, Jason builds data-driven and computational systems that understand, manipulate, and synthesize user interfaces to maximize the usability and accessibility of computers. Jason’s work has received awards at academic conferences and has been by a Fast Company Innovation by Design Student Finalist Award, press coverage in major outlets such as TechCrunch and AppleInsider, and by the FCC Chair Awards for Advancements in Accessibility.

\vspace{2px}

\textbf{Savvas Petridis} (Google DeepMind) is a Research Scientist at Google DeepMind, in the People + AI Research (PAIR) team. He researches Human-AI Interaction, with a focus on human controllability and understanding of large generative models. Recently, his work has focused on helping users formulate and communicate their requirements to these large models and is curious how this process might change over longer periods of usage.

\vspace{2px}

\textbf{Dingzeyu Li} (Adobe) is a Senior Research Scientist at Adobe Research. He is interested in building novel creative tools, using the latest advances from vision, graphics, machine learning, and HCI. His past research and engineering has been recognized by two ACM UIST Best Paper Awards (2022, 2017), an Emmy Award for Technology and Engineering (2020), two Adobe MAX Sneaks Demos (2019, 2020), an Adobe Research Fellowship (2017), a NVIDIA PhD Fellowship Finalist (2017), a Shapeways Educational Grant (2016), and an HKUST academic achievement medal (2013). 

\vspace{2px}

\textbf{Tuhin Chakrabarty} (Salesforce) is a Research Scientist at Salesforce AI Research and an Assistant Professor at Stony Brook University. His research interests are broadly in NLP , Interactivity and Human AI Alignment. He focuses on the design and development of reliable AI systems that can handle implicature, ambiguity, understand human behavior and are aligned with the requirements they have from technology. Tuhin’s work on AI and Creativity has been covered by MIT Technology Review, BloomBerg, Washington Post and The Hollywood Reporter.

\vspace{2px}

\textbf{Yue Jiang} (University of Utah and Aalto University) is an assistant professor at the University of Utah. She graduated from Aalto University and the Finnish Center for AI in Finland. Her research focuses on computational user interface understanding, with specific interests in generating adaptive UIs for different users and contexts, AI-assisted design, and modeling human behavior.

\vspace{2px}

\textbf{Jingyi Li} (Pomona College) is an Assistant Professor of Computer Science at Pomona College, where they direct the Doodle Lab. Their research aims to understand, critique, and construct new computational media authoring tools that are sensitive to the power dynamics inherent in cultural forms of creativity.

\vspace{2px}

\textbf{Tiffany Tseng} (Barnard College) is an Assistant Professor of Computer Science at Barnard College and the director of the Design Tools Lab. Her research contributes to design software that enables creative expression and knowledge sharing practices. Before joining Barnard, she was a research scientist at Apple and Project Assistant Professor at the University of Tokyo. She has developed design tools across creative domains including animation, machine learning, electronics prototyping, and 3D design, both through her research and professional work as a product designer at companies such as Autodesk and IDEO. Her work aims to empower a range of users, from young people to professional designers, to realize their creative potential using new  technologies.

\vspace{2px}

\textbf{Ken Nakagaki} (University of Chicago) is an Assistant Professor at the University of Chicago, directing the Actuated Experience Lab - AxLab. He and his research group explore research in interactive hardware technology, engineering, designing, and speculating actuated user interfaces that 'actuate' people for tangible and embodied interaction.

\vspace{2px}

\textbf{Qian Yang} (Cornell University) is an assistant professor in information science. As a human-AI interaction researcher, she helps translate AI's algorithmic advances into valuable real-world applications that serve human ends. Yang has so far designed several high-consequence AI applications: from decision support systems for life-and-death healthcare decisions to context-aware mobile services, from Natural Language Generation systems to autonomous cars. Building upon this related vein of practice, she works to inform a basic understanding of AI as a material for HCI design, helping integrate AI into people's day-to-day practices. 

\vspace{2px}

\textbf{Nikolas Martelaro} (Carnegie Mellon University) is an Assistant Professor at Carnegie Mellon University. His research focuses on seeks to augment designers' capabilities so that we can best leverage human capacity and computation to solve society's toughest problems. His work spans design domains, blending hardware, software, and interaction design.

\vspace{2px}

\textbf{Jeffrey V. Nickerson} (Stevens Institute of Technology) is the Steven Shulman ’62 Chair for Business Leadership and Professor of Digital Innovation at Stevens Institute of Technology. His research focuses on how humans and machines work together in creative endeavors such as design. He is currently an investigator in a National Science Foundation funded project called the Future of News Work, which is looking at the effects of generative AI on journalism and related fields. 

\vspace{2px}

\textbf{Lydia B. Chilton} (Columbia University) is an Associate Professor in the Computer
Science Department at Columbia University. Her research is in
computational design - how computation and AI can help people
with design, innovation, and creative problem-solving. Applications
include: creating media for journalism, developing technology for
public libraries, improving risk communication during hurricanes, 
helping scientists explain their work, and improving mental health
in marginalized communities.

\end{document}